\documentclass[12pt]{iopart}

\usepackage{iopams}
\expandafter\let\csname equation*\endcsname\relax
\expandafter\let\csname endequation*\endcsname\relax
\usepackage{amsmath}  % needed for \tfrac, \bmatrix, etc.
\usepackage{amsfonts} % needed for bold Greek, Fraktur, and blackboard bold
\usepackage{siunitx} %\SI[separate-uncertainty = true]{1.0(1)e-3}{\second}
\usepackage{verbatim}
\usepackage{color}
\usepackage[english]{babel}
\usepackage{blindtext}
\usepackage[pdftex]{graphicx}
\usepackage{epstopdf} 
\usepackage{hyperref}
\usepackage{cite}
\usepackage[version=4]{mhchem}

\begin{document}

\title{Automation of the Cavendish Experiment to `Weigh the Earth'}
% In a long title you can use \\ to force a line break at a certain location.

\author{J J Andrews and J S Bobowski}

\address{Department of Physics, University of British Columbia, Kelowna, British Columbia, Canada V1V 1V7}
\ead{jake.bobowski@ubc.ca}

\begin{abstract}
We describe a simple and inexpensive method for automating the data collection in the well-known Cavendish torsion balance experiment to determine the gravitational constant $G$.  The method uses a linear array of phototransistors and requires no moving parts.  Multiplexers and a data-acquisition device are used to sample the state of each phototransistor sequentially.  If the sampled phototransistor is illuminated by the laser spot, the position and time are recorded to a data file.  The recorded data does an excellent job of capturing the damped harmonic oscillations.  The resulting data were analysed to extract an experimental value of $G$ that was within 5\% of the accepted value.       

~


~

\noindent{(Some figures may appear in colour only in the online journal)\/}
\end{abstract}

%\keywords{gravitational constant, automation, torsion balance}
%\submitto{\EJP}
%\maketitle
%\ioptwocol % two column

\section{Introduction}\label{sec:intro}
Using a torsion balance to measure the gravitational constant $G$ has long been one of the standard experiments offered in undergraduate physics programs.  The apparatus beautifully eliminates the effects of the gravitational force on the torsion balance due to the Earth (and all other nearby stationary objects)~\cite{Cavendish:1798}.  In the experiment, students disturb the equilibrium of the torsion balance by changing the positions of a pair of \SI{1.5}{\kilo\gram} lead spheres and observe the system as it approaches the new equilibrium.  Because the system is underdamped, the torsion balance oscillates about the new equilibrium position with period $T$ and a decay time constant $\tau\gg T$.  In a typical apparatus, the force of the large lead spheres on the torsion balance is of order \SI{1}{\nano\newton} which highlights the sensitivity of the method.

Once the apparatus is properly aligned, the experiment requires students to collect data continuously for at least \SI{90}{\minute} and, during that time, vibrations in the room must be kept to a minimum.  In this paper we describe a simple and relatively inexpensive method to automate the data collection.  Our method allows students to disturb the equilibrium of the torsion balance, start the data acquisition program, and then leave the room.  After approximately \SI{1}{\hour}, students return to move the large lead spheres back to their original positions without stopping the data acquisition.  Our system acquires higher resolution data at a higher sampling rate than is possible manually.  Furthermore, because students are only required to change the positions of the lead spheres approximately once per hour, they typically are able to run additional trials which allows for a better measurement of the shift in equilibrium position.

Of course, others have already describe methods for automating the Cavendish torsion balance experiment~\cite{Fischer:1987, Fitch:2007, Bach:2007, Tomarken:2012}.  One method of automation involves mounting a pair of photodetectors on movable translation stage.  The photodetectors are positioned close to one another such that they can be simultaneously illuminated by a laser beam reflected from a mirror mounted on the torsion balance.  Electronics are used to compare the outputs of the two photodetectors.  If the detector on the right is collecting more light, the cart moves right until the detector outputs become equal.  Recording the position of the cart as a function of time allows one to accurately measure the damped oscillations of the torsion pendulum~\cite{Fischer:1987, Fitch:2007}.  A second automation method that has been used is to analyse a video recording of the reflected laser beam to track its position as a function of time.  The simplest method is to record the video and then process it afterwards using an image-analysis software package.  Other methods have been developed to extract the required data from the video in real time~\cite{Bach:2007, Tomarken:2012}.  Rather than track the position of a reflected laser spot, TEL-Atomic, Inc.\ manufactures a torsion balance that has a built-in capacitive sensor that determines the angular displacement of the suspended beam~\cite{TEL-Atomic}.  It is also worth noting that others have used stepper motors to automate the process of changing the positions of the large external masses~\cite{Thompson:2000}.  The advantages of the method that we have developed are that it requires only inexpensive components, has no moving parts and is therefore simple and robust, and does not require any complex algorithms to extract meaningful data.

Finally, we point out that, despite the fact that the Cavendish results were published in 1798~\cite{Cavendish:1798}, sensitive torsion balances have continued to be used in modern physics research and will continue to be used for the foreseeable future~\cite{Beth:1936, Kapner:2007, Adelberger:2009, Kim:2016, Ahn:2018, Balushi:2018}.  Therefore, the skills and experience that students gain from performing the classic Cavendish ``experiment to determine the density of the Earth'' undoubtedly remain relevant.    

\section{Theory}

\begin{figure}[t]
\centering{
\small{(a)}\includegraphics[keepaspectratio, width=0.40\columnwidth]{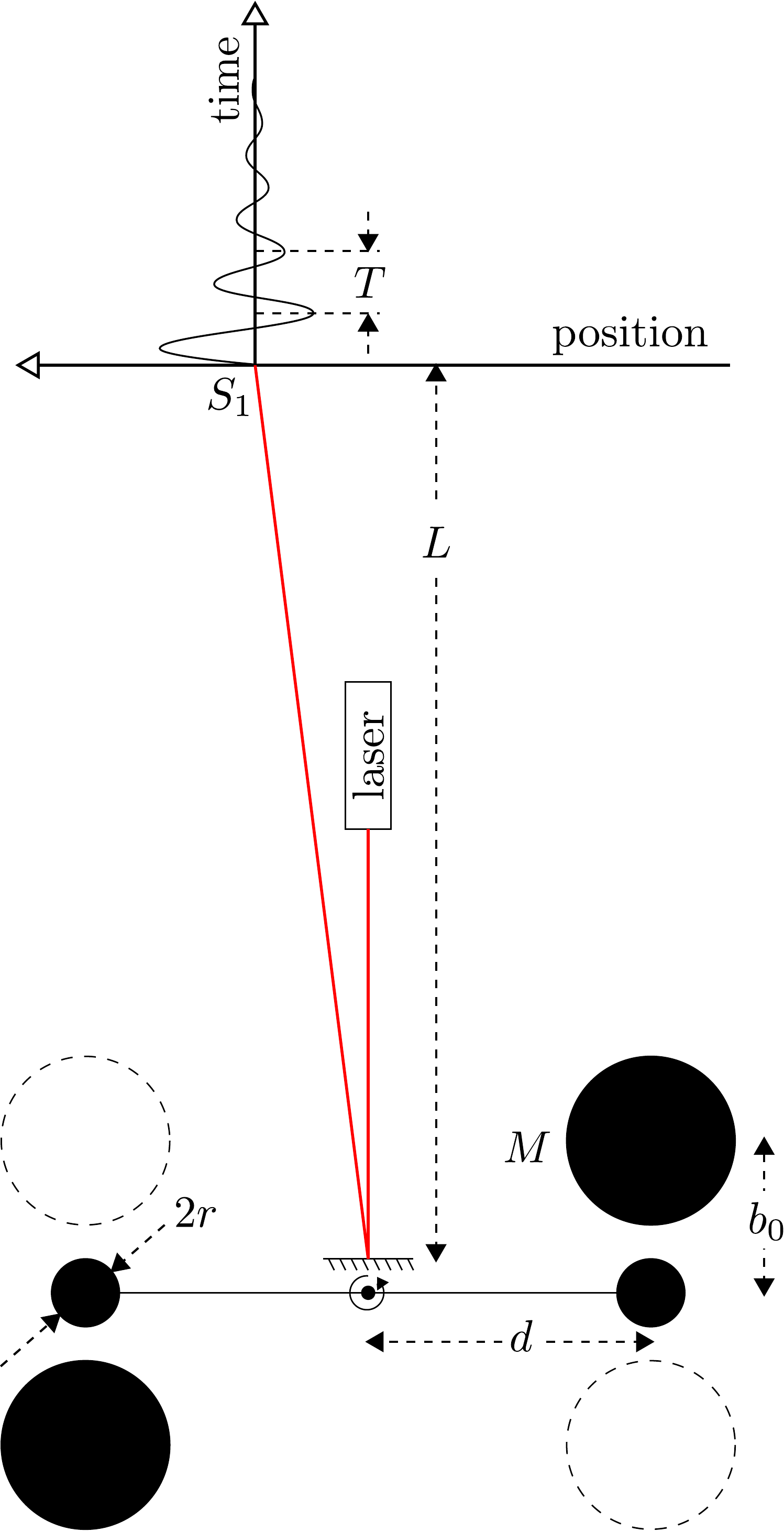}\qquad \small{(b)}\includegraphics[keepaspectratio, width=0.46\columnwidth]{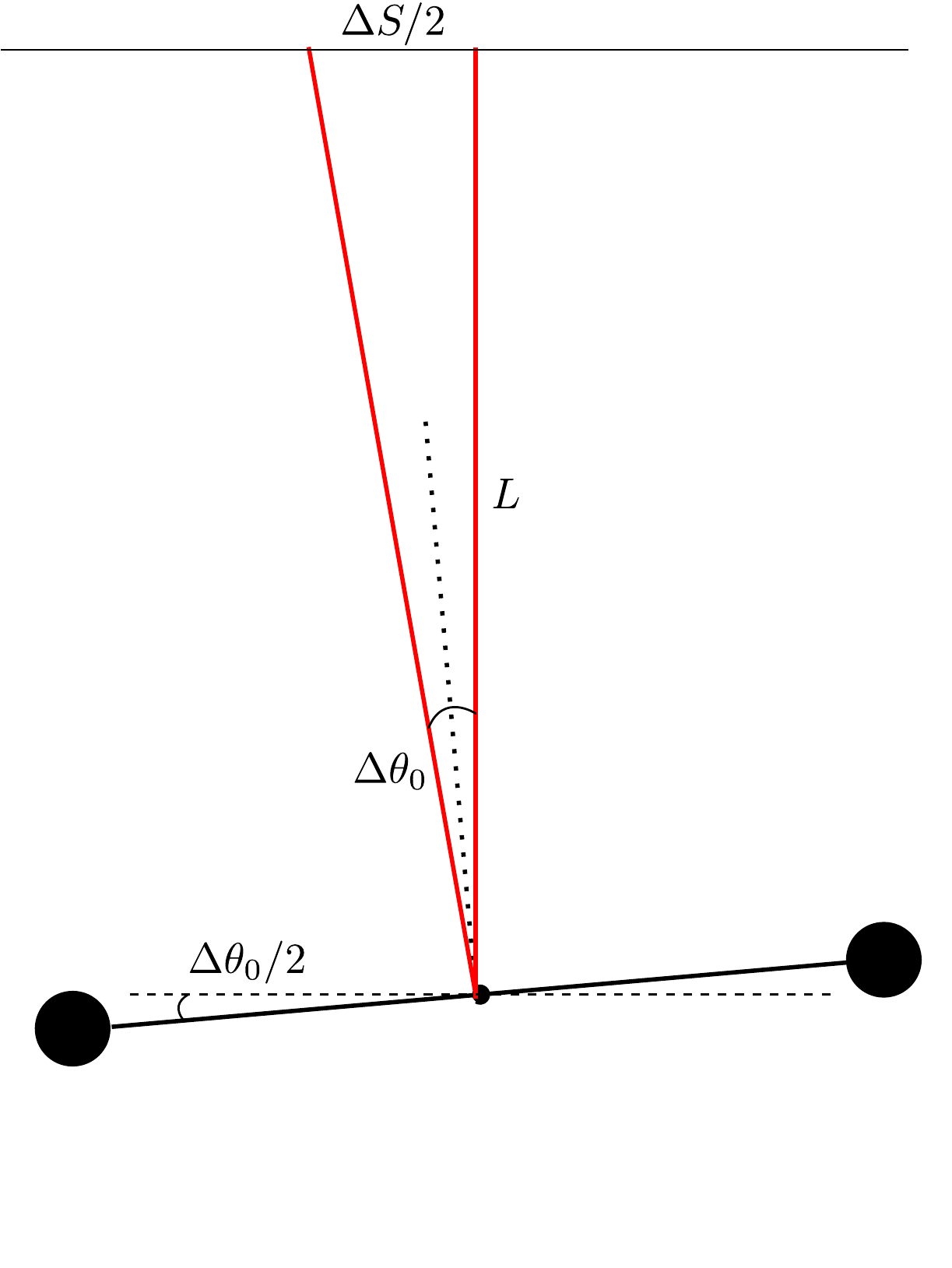}}
\caption{\label{fig:geo}(a) Experimental geometry for the Cavendish experiment.  A laser is directed towards a mirror mounted to the torsion balance.  The reflected spot oscillates with period $T=2\pi/\omega_1$ about an equilibrium position $S$.  The figure shows the large masses $M$ in position ``1'' such that the equilibrium position is denoted $S_1$.  When the large masses are moved to position ``2'' (indicated by dashed circles), the equilibrium will shift to a new position ($S_2$).    (b) Geometry used to relate the change in the angular equilibrium position $\Delta\theta_0$ to the measured linear displacement $\Delta S$. }
\end{figure}

The gravitational force on one of the small lead sphere (mass $m$) due to a nearby large lead sphere (mass $M$) is given by $GMm/b_0^2$ where $G$ is the gravitational constant and $b_0$ is the centre-to-centre distance between the pair of spheres. See Fig.~\ref{fig:geo}(a) for a schematic of the experimental geometry.  
As the torsion pendulum twists away from its equilibrium position, the magnitude of the restoring torque provided by the torsion ribbon is $\kappa\theta$ where $\kappa$ is a constant and $\theta$ is the angular displacement away from equilibrium.  If a drag torque proportional to $\dot\theta$ is assumed, then the net torque on the torsion balance at any instant in time is given by:
\begin{equation}
I\ddot\theta = 2\frac{GMm}{b_0^2}d-\frac{2I}{\tau}\dot\theta-\kappa\theta,\label{eq:diff}
\end{equation}
where $I$ is the moment of inertia of the pendulum, $d$ is the distance from the centre of the small sphere to the rotation axis of the pendulum, and $\tau$ is the damping time constant.  A factor of two has been inserted in front of the first term on the right-hand side because there are two sets of identical small and large lead spheres.  We assume that, as the small spheres oscillate, the value of $b_0$ remains approximately constant.  

Setting $\dot\theta=\ddot\theta=0$ gives the equilibrium value of $\theta$:
\begin{equation}
\theta_0=\frac{2}{\kappa}\frac{GMm}{b_0^2}d,\label{eq:equil}
\end{equation}
such that Eq.~(\ref{eq:diff}) can be written:
\begin{equation}
\omega_0^2\,\theta_0=\ddot\theta+\frac{2}{\tau}\dot\theta+\omega_0^2\,\theta,\label{eq:diff2}
\end{equation}
where $\omega_0^2\equiv \kappa/I$.  In the underdamped case, $\omega_0> 2/\tau$, the solution to Eq.~(\ref{eq:diff2}) is:
\begin{equation}
\theta(t)=\theta_0-\theta_A e^{-t/\tau}\cos\left(\omega_1 t\right),\label{eq:soln}
\end{equation}
where $\omega_1^2=\omega_0^2-\tau^{-2}$ and the initial amplitude of the oscillations $\theta_A$ is determined from the initial conditions.  The observed oscillation frequency of the torsion balance is given by $\omega_1$ while $\omega_0$ represents the oscillation frequency in the zero-damping limit ($\tau\to\infty$).  Measurements of $\omega_1$ and $\tau$ can be used to determine the torsion constant $\kappa$ of the fibre:
\begin{equation}
\frac{\kappa}{I}=\omega_1^2+\frac{1}{\tau^2}.\label{eq:kappa}
\end{equation}
For highly-underdamped oscillations $\left(\omega_1\approx\omega_0\gg\tau^{-1}\right)$, $\kappa\approx I\omega_1^2$.

The equilibrium angle $\theta_0$ given by Eq.~(\ref{eq:equil}) is valid for one position of the large lead spheres.  If the spheres are moved to the second position (dashed circles in Fig.~\ref{fig:geo}(a)), by symmetry, the new equilibrium will be $-\theta_0$ such that the change in the equilibrium angle is:
\begin{equation}
\Delta\theta_0=2\theta_0=\frac{4}{\kappa}\frac{GMm}{b_0^2}d.\label{eq:dtheta}
\end{equation}
Solving for $G$ and using Eq.~(\ref{eq:kappa}) to eliminate $\kappa$ yields:
\begin{equation}
G=\frac{\Delta\theta_0 b_0^2I}{4Mmd}\left(\omega_1^2+\frac{1}{\tau^2}\right).
\end{equation}
As Fig.~\ref{fig:geo}(b) shows, the angular displacement $\Delta\theta_0$ can be expressed in terms of the linear displacement of the reflected laser spot.  The linear displacement $\Delta S/2$ is, to a very good approximation, equal to the arclength $L\Delta\theta_0$, where $L$ is the distance from the mirror to the screen. Finally, if the small lead spheres have radius $r$, moment of inertia of the torsion pendulum is \mbox{$I=2m\left(d^2+2r^2/5\right)$} such that:
\begin{equation}
G=\frac{\Delta S\, b_0^2}{4MdL}\left(d^2+\dfrac{2}{5}r^2\right)\left(\omega_1^2+\frac{1}{\tau^2}\right).\label{eq:G}
\end{equation}
All of the variables in Eq.~(\ref{eq:G}) are measurable such that the gravitational constant $G$ can be experimentally determined.

\section{Automation method}\label{sec:automation}
\begin{figure}[t]
\centering{
\includegraphics[keepaspectratio, width=0.65\columnwidth]{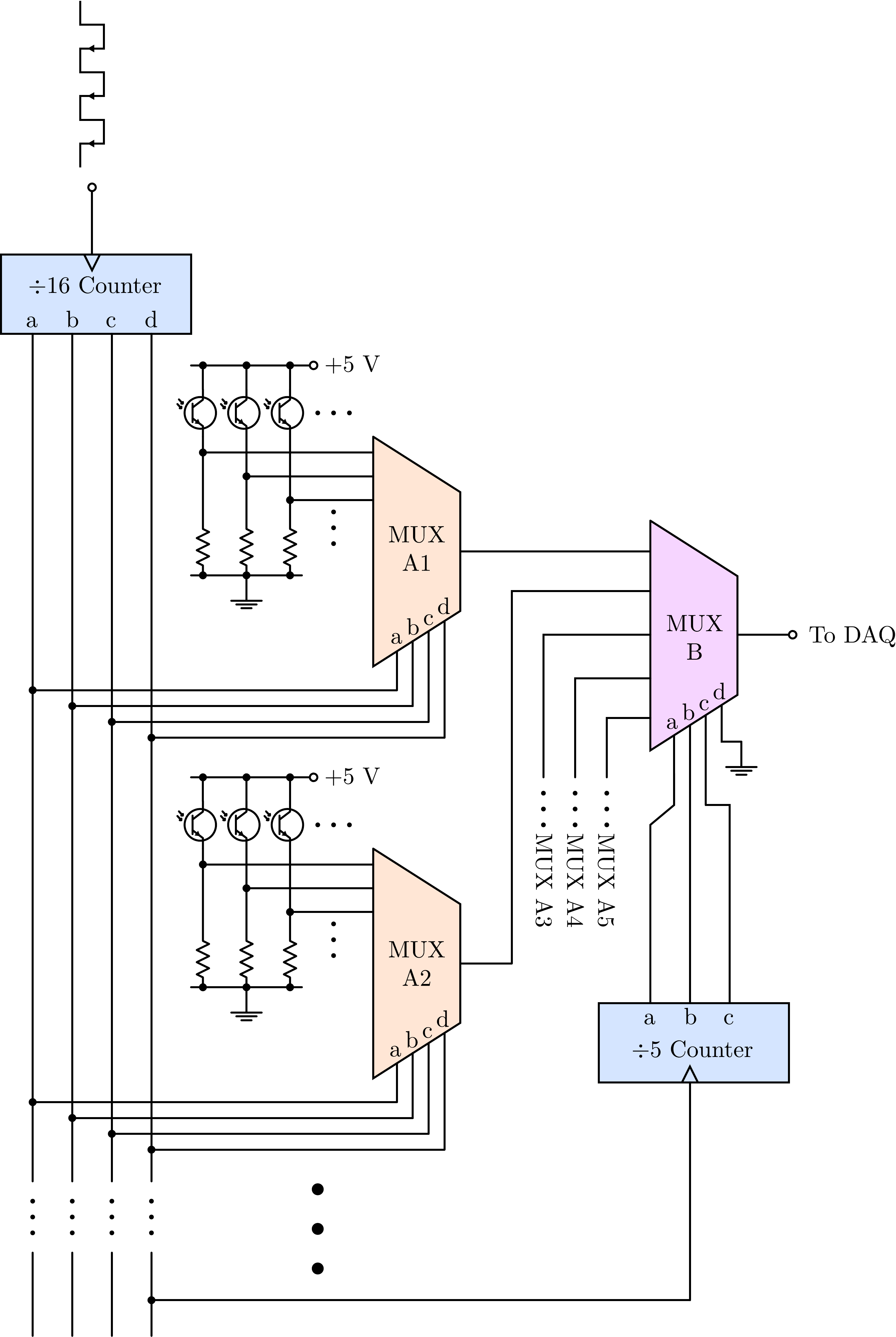}}
\caption{\label{fig:MUX}The circuit developed to sequentially sample the states of the phototransistors in the array.  Multiplexers A1 through A5 (only A1 and A2 are shown) are connected directly to the phototransistors.  Each of the ``A'' MUXs have 16 inputs, although only three are shown for clarity.  The outputs of the A MUXs are fed into MUX B and its output is monitored by a DAQ.}
\end{figure}

To automate the data collection, a linear array of 80 NPN phototransistors was constructed.  As shown in Fig.~\ref{fig:MUX}, each phototransistor in the array had its collector connected to \SI{5}{\volt} and its emitter was connected to ground via a \SI{1}{\kilo\ohm} resistor. 
The state (conducting or not conducting) of the phototransistor was monitored via the voltage across the \SI{1}{\kilo\ohm} resistor.  When the phototransistor is not illuminated by the laser spot, the current from collector to emitter is zero and the voltage across the resistor is LO.  When the phototransistor is illuminated, there is a nonzero current and the voltage across the resistor is read as HI.  

To sequentially check the state of each phototransistor in the array, a relatively simple circuit consisting of multiplexers (MUX) and counters was developed.  Five 16-input binary MUXs were used to monitor the states of the 80 phototransistors.  As shown in Fig.~\ref{fig:MUX}, MUX A1 was used to monitor phototransistors 1 to 16, A2 to monitor 17 - 32, and so on.  The 4-bit output of a single $\div 16$ counter was used to select the address of each MUX A1 to A5.  If, for example, the selected address is $0011$, then A1 outputs the state of detector 4, A2 outputs the state of detector 20,\dots and A5 outputs the state of detector 68.  On the negative edge of the next clock pulse, the address will increment to $0100$ and the states of detectors 5, 21, 37, 53, and 69 will be passed to the outputs of the ``A'' MUXs.  These five outputs are passed to MUX B whose address is controlled by a $\div 5$ counter.  The $\div 5$ counter increments only after the $\div 16$ has completed an entire sequence of counts and resets to zero.  In this way, the first cycle of 16 counts is used to check the states of detectors 1 to 16, the second cycle checks the states of detectors 17 to the 32, and so on.  The final output of MUX B is passed to a data acquisition device (DAQ) that is controlled by a simple LabVIEW program.  

The DAQ that we used to control the circuit and acquire the data was the National Instruments USB-6001.  However, any DAQ with two digital outputs and one digital input will work.  When the data-acquisition program is started, one of the digital outputs is used to reset both counters so that the starting count is zero.  Next, the digital input reads the output of MUX B (LO or HI) which corresponds to the state of the first detector in the phototransistor array.  If the reading is HI, then the known position of the detector and the time elapsed since the program was started are written to a data file.  If the reading is LO, then no data is recorded.  Next, the second digital output of the DAQ is cycled from LO to HI and back to LO which advances the count of the $\div 16$ counter so that the state of the next detector in the array can be tested.  This process continues until the user stops to program.  When the $\div 16$ counter reaches a count of 15 (1111, in binary), it will reset to zero on the next clock pulse and the $\div 5$ counter will increment by one.

\begin{figure}[t]
\centering{
\includegraphics[keepaspectratio, width=0.65\columnwidth]{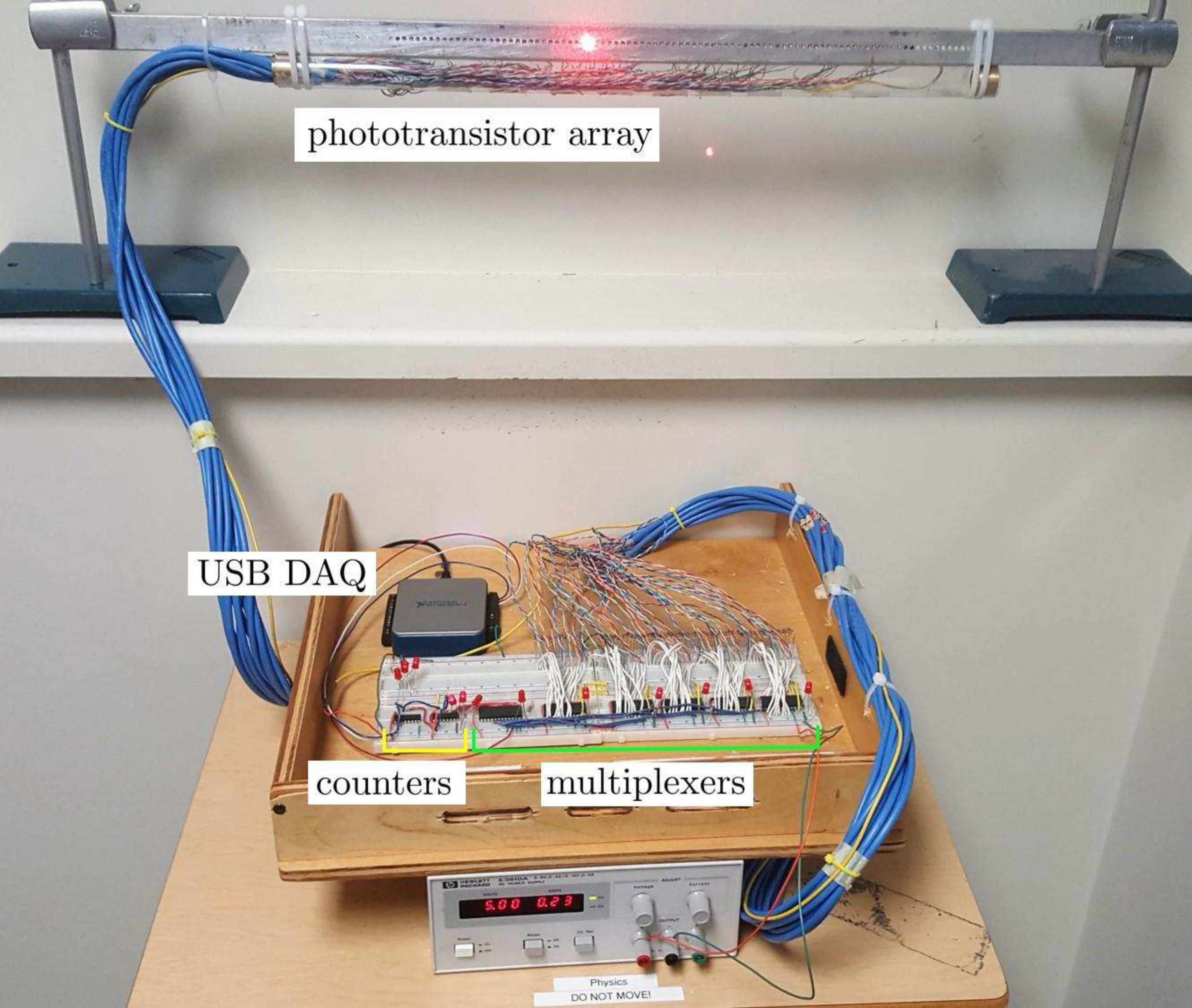}}
\caption{\label{fig:photo}Photograph of the phototransistor array, multiplexer circuit, and data acquisition (DAQ) device. The photograph exaggerates the size of the laser spot.  Its size is approximately \SI{3}{\milli\meter} in diameter and it illuminates only a single phototransistor at a time.}
\end{figure}

A photograph of the circuit and detector array is shown in Fig.~\ref{fig:photo}. 
The Vishay BPW85 phototransistors that we used have a diameter of \SI{3}{\milli\meter} and they were mounted in a long aluminum U-channel.  A series of \SI[number-unit-product=\text{-}]{3}{\milli\meter} diameter holes were drilled into the aluminum.  The centre-to-centre spacing between adjacent holes was \SI{5}{\milli\meter}.  The phototransistors were held in place using a small amount of epoxy.  We used SN74150N multiplexers made by Texas Instruments and two Motorola 74HC390 counters.  Each counter chip has two $\div 2$ counters and two $\div 5$ counters.  The four $\div 2$ counters were combined to form the required $\div 16$ counter shown in Fig.~\ref{fig:MUX}.  The total cost of the required circuit components was approximately \SI{50}[\$]{USD} and the cost of the DAQ was \SI{210}[\$]{USD}.  

\section{Experimental results and discussion}\label{sec:expt}
If the equilibrium of the torsion pendulum is disturbed by moving the large masses from position 1 to position 2 in Fig.~\ref{fig:geo}(a), the initial amplitude of the resulting oscillation of the laser spot will be \mbox{$\Delta S=S_2-S_1$}.  The observed value of $\Delta S$ using our experimental setup, see table~\ref{tab:param}, was \SI{1.87}{\centi\meter}.  An oscillation of this size would only sample three of the phototransistors in our array.   
\begin{table}
\caption{\label{tab:param}Measured parameter values used to determine the gravitational constant $G$.  The values of $\omega_1$, $\tau$, and $\Delta S$ were obtained from fits to the data shown in Fig.~\ref{fig:data}.}
\begin{indented}
\item[]\begin{tabular}{rl}
\br
parameter & measurement\\
\hline\hline
$r$ & \SI[separate-uncertainty = true, multi-part-units=single]{7.56\pm 0.05}{\milli\meter}\\
$b_0$ & \SI[separate-uncertainty = true, multi-part-units=single]{46.4\pm 0.1}{\milli\meter}\\
$d$ & \SI[separate-uncertainty = true, multi-part-units=single]{49.9\pm 0.5}{\milli\meter}\\
$L$ & \SI[separate-uncertainty = true, multi-part-units=single]{243.5\pm 1.0}{\centi\meter}\\
$M$ & \SI[separate-uncertainty = true, multi-part-units=single]{1500\pm 1}{\gram}\\
\hline
$\omega_1$ & \SI[separate-uncertainty = true, multi-part-units=single]{0.0216111\pm 0.0000012}{\radian/\second}\\
$\tau$ & \SI[separate-uncertainty = true, multi-part-units=single]{1068.6\pm 1.4}{\second}\\
$\Delta S$ & \SI[separate-uncertainty = true, multi-part-units=single]{1.87\pm 0.14}{\centi\meter}\\
\br
\end{tabular}
\end{indented}
\end{table}
In order to have the laser spot sample the majority of the photodetectors, which span \SI{39.5}{\centi\meter}, it is necessary to start with larger amplitude oscillations.

One way to build up the oscillation amplitude is to change the position of the large spheres each time the torsion pendulum reaches an end point of its swing.  Once the desired amplitude has been achieved, the large masses are left stationary and one begins collecting data~\cite{Thompson:2000}.  The disadvantages of this method are that (1) it requires some time to execute since the period of oscillations is approximately \SI{5}{\minute} and (2) each time the positions of the large masses are changed, one risks accidentally bumping the sensitive apparatus and causing unwanted vibrations which can take \SI{20}{\minute} or more to dissipate.

We have found it convenient to use a strong rare-earth magnet to quickly induce large-amplitude oscillations of the pendulum without making any physical contact with the apparatus.  With the pendulum at rest and the large masses in the desired position, we {\it slowly} moved a \ce{Nd2Fe14B} disk magnet near one of the small masses $m$ of the pendulum.  The magnet was then quickly pulled away in a direction transverse to the plane of oscillations.  The quick motion causes a non-negligible change in magnetic flux $d\Phi/dt$ that induces a current in the small lead sphere.  The magnetic moment created by the induced current causes the sphere to be attracted to the magnet which triggers the oscillations.  It is relatively easy to manipulate the initial amplitude of the oscillations by controlling the speed with which the magnet is pulled away.  In our experiments, we used a \SI[number-unit-product={\text{-}}]{25}{\milli\meter} diameter and \SI[number-unit-product={\text{-}}]{12.5}{\milli\meter} thick \ce{Nd2Fe14B} disk magnet.      

In Eq.~(\ref{eq:diff}), we assumed that the centre-to-centre distance $b_0$ between the large and small masses remains approximately constant during the oscillations.  One may legitimately worry that we violate this assumption when we induce large-amplitude oscillations using the magnet.  In our apparatus, $b_0$ can vary by up to a maximum of than 10\%.  It is important to emphasize, however, that small variations in $b_0$ have only a relatively minor affect on the average torque that the large spheres apply to the pendulum.  An oscillation of the pendulum will cause the centre-to-centre distance between the lead spheres to vary as:
\begin{equation}
b(t)=b_0+\delta\sin\left(\omega_1 t\right),
\end{equation}
where $\delta\ll b_0$.  The resulting torque, averaged over one period, is greater than the equilibrium torque by a factor of $\approx 1+(3/2)\left(\delta/b_0\right)^2$.  Therefore, a 10\% variation of $b_0$ will increase the average torque by only 1.5\%.  Furthermore, $\delta/b_0$ is typically much less than 0.1.

\begin{figure}[p]
\centering{
\includegraphics[keepaspectratio, width=0.7\columnwidth]{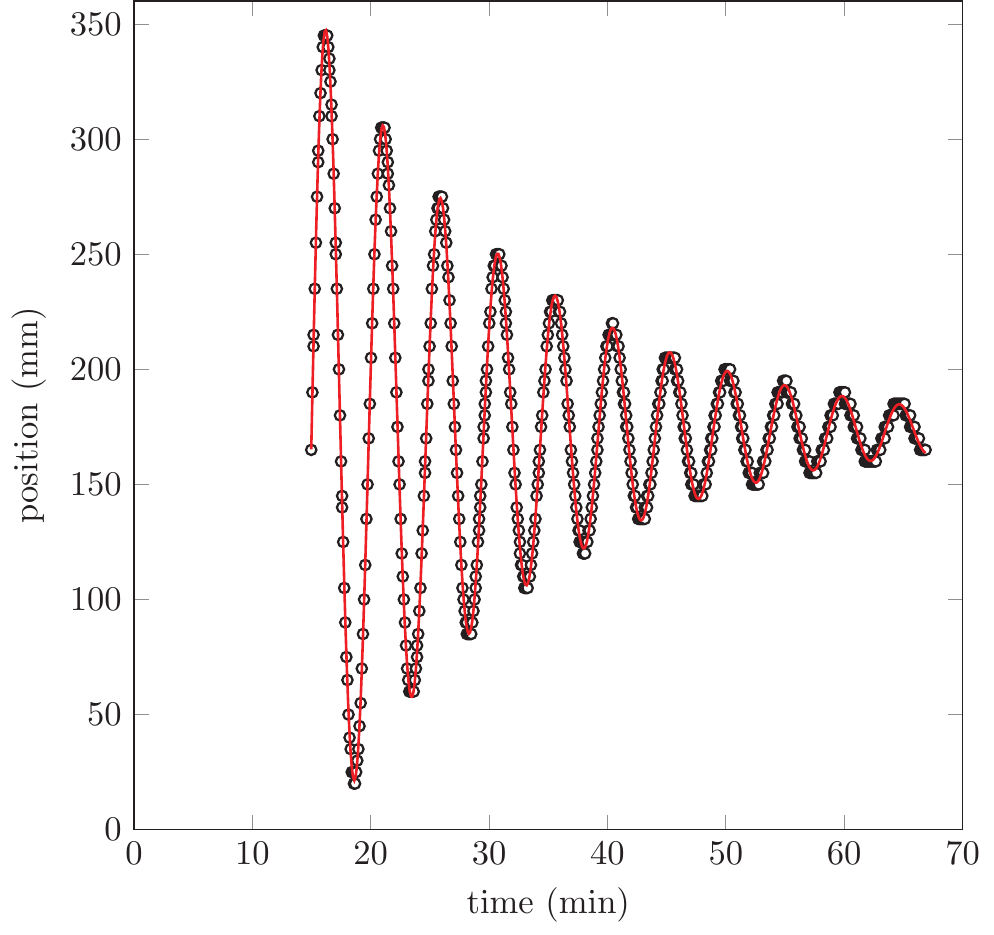}}
\caption{\label{fig:dataShort}Laser spot position measured as a function of time (black circles) fit to a damped harmonic oscillation (red curve).}
\end{figure}

\begin{figure}[p]
\centering{
\includegraphics[keepaspectratio, width=\columnwidth]{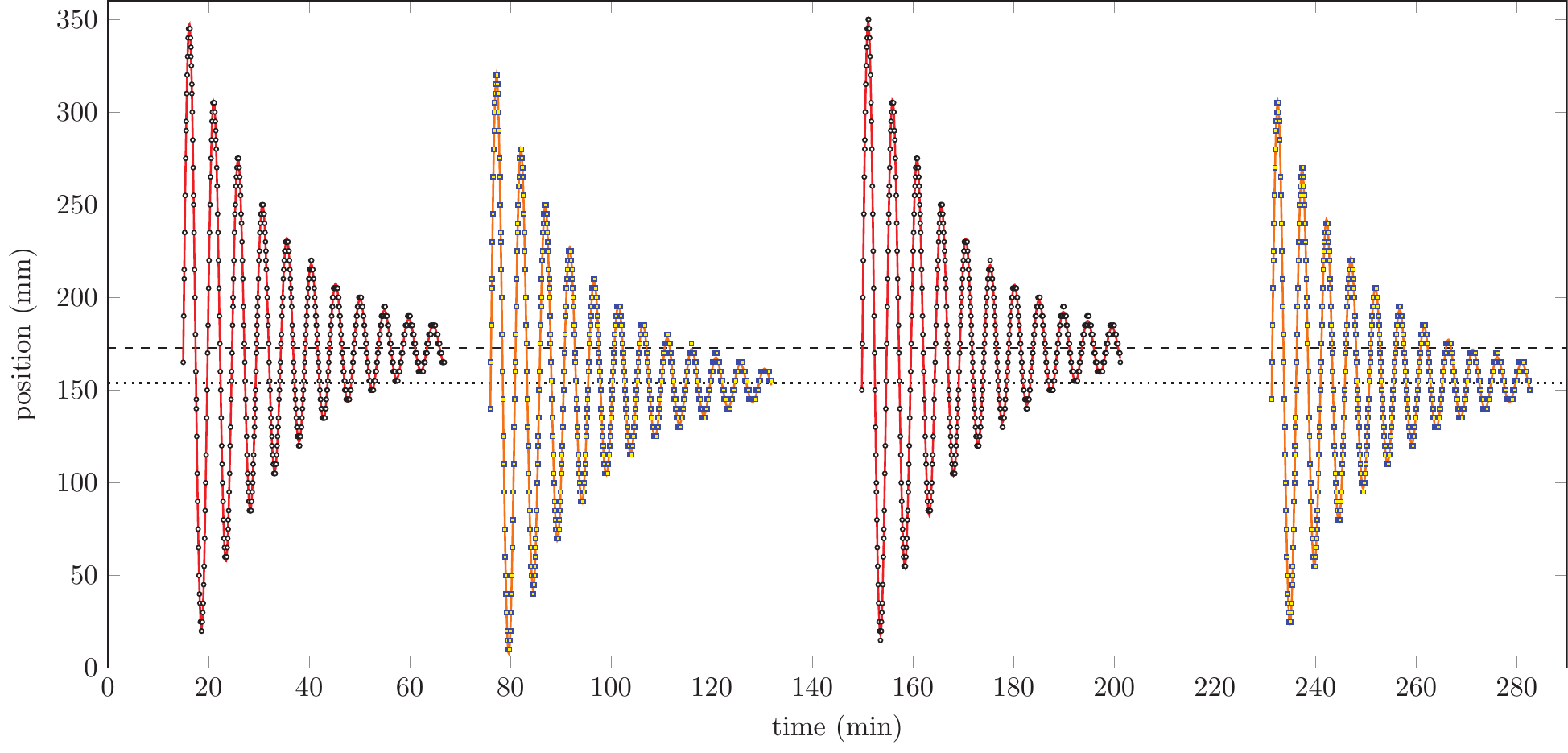}}
\caption{\label{fig:data}Position of the reflected laser versus time with the large lead spheres in position 1 (black circles) and position 2 (blue squares).  The change in equilibrium position from $S_1$ (dashed line) to $S_2$ (dotted line) is clearly visible.}
\end{figure}

Figure~\ref{fig:dataShort} shows one set of damped oscillations recorded using our automated data-acquisition system.  The first cycle of the oscillations spans approximately \SI{380}{\milli\meter} which is 95\% of the length of our detector array.  The DAQ device is able to sequentially check the state of all 80 phototransistors in \SI{4.4}{seconds} which corresponds to a sampling rate of \SI{18}{\hertz}.  The red line in Fig.~\ref{fig:dataShort} is a fit to the data using a function of the form given in Eq.~(\ref{eq:soln}).  The fit is excellent and gives reliable values for $\omega_1$, $\tau$, and the equilibrium position.

Figure~\ref{fig:data} shows a series of four damped oscillations measured using our system, two each for the masses in position 1 and 2.  The first set of oscillations are identical to the data shown in Fig.~\ref{fig:dataShort}.  In all four cases, the data have been fit to extract $\omega_1$, $\tau$, and the equilibrium position.  The weighted means of the $\omega_1$ and $\tau$ values are given in table~\ref{tab:param}.  Notice that the relative error in $\Delta S$ is \SI{0.075}{} which is much greater than the relative error in any of the other measure quantities.  Since, as shown in Eq.~(\ref{eq:G}), \mbox{$G\propto \Delta S$}, we expect the uncertainty in our experimental value for $G$ to be close to 7.5\%.  When the results from table~\ref{tab:param} are inserted to Eq.~(\ref{eq:G}), we obtain \mbox{$G=\SI[separate-uncertainty = true]{6.5(5)e-11}{\meter^3\per\kilo\gram^{-1}\second^{-2}}$}.  This result is in agreement with the expected value of \SI{6.67e-11}{\meter^3\kilo\gram^{-1}\second^{-2}}.  The uncertainty in our experimental value for $G$ was calculated using standard propagation of error methods. 

\begin{comment}
It is common to correct the measured value of $G$ by taking into account the torque exerted by the large masses on the small masses at the far end of the pendulum~\cite{Fischer:1987, Fitch:2007}.  The corrected value of the gravitational constant is greater than the uncorrected value and is given by \mbox{$G_0=G\left(1-B\right)^{-1}$}, where:
\begin{equation}
B=\frac{b_0^3}{\left(b_0^2+4d^2\right)^{3/2}}.
\end{equation}
When we apply this correction to our data we get a value of $G_0$ that is greater than the expected value, but still in agreement within uncertainty. 
\end{comment}

\section{Summary}
We have described a simple and inexpensive multiplexer circuit and phototransistor array that can be used to automate the data collection in the classic torsion balance experiment to measure the gravitational constant.  The system is low maintenance and can be left unattended for long periods of time to acquire data at a reasonably high sampling rate.  The data acquisition system works best with relatively large-amplitude oscillations which are easy to induce using a strong permanent magnet.  The large amplitude-oscillations do not compromise the quality of data in any serious way.  Using this system, students can easily acquire multiple trials of high-quality data in a way that is not onerous.  When using the automated data collection, we routinely obtain experimental values for $G$ that are within 5\% of the expected value.

\end{document}